\documentclass[proof]{WileyASNA-v1}

\articletype{Article Type}%

\received{\today}
\revised{\today}
\accepted{\today}

\begin{document}

\title{Quantifying the impact of relativistic precession on tidal disruption event light curves}

\author[1]{Diego Calderón*}

\author[2]{Ondřej Pejcha}

\author[3,4]{Brian D. Metzger}

\author[5]{Paul C. Duffell}

\author[6,7]{Stephan Rosswog}

\authormark{Calderón \textsc{et al}}

\address[1]{\orgname{Max-Planck-Institut für Astrophysik}, \orgaddress{Karl-Schwarzschild-Straße 1, 85748 \state{Garching}, \country{Germany}}}

\address[2]{\orgdiv{Institute of Theoretical Physics, Faculty of Mathematics and Physics}, \orgname{Charles University}, \orgaddress{V Holešovičkách 2, 180 00 \state{Prague}, \country{Czech Republic}}}

\address[3]{\orgdiv{Department of Physics and Columbia Astrophysics Laboratory}, \orgname{Columbia University}, \orgaddress{\state{New York, NY 10027}, \country{USA}}}

\address[4]{\orgdiv{Center for Computational Astrophysics}, \orgname{Flatiron Institute}, \orgaddress{\state{162 5th Avenue, New York, NY 10010}, \country{USA}}}

\address[5]{\orgdiv{Department of Physics and Astronomy}, \orgname{Purdue University}, \orgaddress{525 Northwestern Avenue,  West Lafayette, \state{IN 47907-2036}, \country{USA}}}

\address[6]{\orgdiv{Hamburger Sternwarte}, \orgname{Universit\"at Hamburg}, \orgaddress{Gojenbergsweg 112}, \state{21029 Hamburg}, \country{Germany}}

\address[7]{\orgdiv{The Oskar Klein Centre, Department of Astronomy, AlbaNova}, \orgname{Stockholm University}, \state{SE-106 91 Stockholm}, \country{Sweden}}

\corres{*Diego Calderón. \\ \email{calderon@mpa-garching.mpg.de}}

\presentaddress{
    Karl-Schwarzschild-Straße 1
    \\
    85748 Garching bei München, Germany
    }

\abstract{
    The tidal field of a black hole can turn a star into a gas stream whose orbit can precess, especially if the a black hole is rapidly spinning. 
    In this work, we investigate the impact of precession on the light curves of tidal disruption events (TDE). 
    To do so, we perform two-dimensional radiation-hydrodynamic simulations of the interaction of the TDE wind and luminosity with the precessed stream wrapped around the black hole. 
    Our results show that in events with black holes of $\sim10^6~\text{M}_{\odot}$ and no orbit-spin inclination, the line of sight has little effect on the light curves, since the stream covers a small fraction of the solid angle as the precession is confined to the orbital plane.
    In the case of black holes of $\gtrsim10^7~\text{M}_{\odot}$ and high inclination ($i\sim90^{\circ}$), the light curve peaks can be delayed by $\sim$100 days due to presence of the precessed stream blocking the radiation in the early phase of the event.
    We also discuss our efforts to model self-consistently the hydrodynamic evolution of a tidal stellar stream on curved spacetimes by the presence of a massive black hole.
}



\fundingInfo{Funding info text.}

\maketitle


\section{Introduction}
\label{sec:intro}

    Super-massive black holes (SMBHs) are thought to be ubiquitous in the centre of galaxies \citep[e.g.][]{kormendy2013}. 
    Since stellar density is higher towards galactic centres some stars pass close enough to the SMBH so that they result in a tidal disruption event (TDE). 
    In the generic scenario, this occurs when the tidal field of the SMBH overcomes the self-gravity of the star which is stretched into a stellar stream that proceeds to wrap around the black hole. 
    Once the stream collides with itself the event is triggered and, as a consequence, a fraction of the material forms an accretion disc and an electromagnetic signature is generated \citep{hills1975,rees1988}. 
    The TDE light curve is characterised by a rapid rise to a peak luminosity followed by a decay with a characteristic power law $\propto t^{-5/3}$, provided the star was fully disrupted \citep[e.g.][]{gezari2006}. 

    The outcome of a TDE due to the action of a rapidly spinning black hole might be significantly different, especially if the incoming star initially moves on an misaligned orbit with respect to the black hole spin \citep{guillochon2015}. 
    This scenario might result into the stellar stream missing its self-collision on its first pericentre passage \citep{stone2012,dai2013}. 
    If so, the stream could wrap around the black hole many more times depending on the properties of the event, i.e. stellar mass, black hole mass and spin, misalignment, and pericentre distance. 
    In this context, the TDE triggering might be delayed and, more importantly, the initial conditions of the event will be different due to the presence of the leftover of the stellar stream around the black hole. 
    Thus, these light curves could show different behaviours as the radiation and outflow of the event will unavoidably interact with the ``ball of yarn" that the stellar stream has become. 
    In this work, we present radiation hydrodynamic calculations of this outcome and make predictions on how this scenario differs from the fiducial picture. 
    Additionally, we discuss our current efforts to estimate more realistic initial conditions for the simulations based on self-consistent relativistic hydrodynamic models. 

\section{Numerical simulations}
\label{sec:sims}

    \subsection{The ``ball of yarn" scenario}
    \label{sec:model}

        Let us consider a star of mass $m_*$ and radius $R_*$, and a black hole of mass $M_\text{h}$ and (dimensionless) spin $a_\text{h}$. 
        If such a star travels inside the tidal radius $r_\text{t}=(M_\text{h}/m_*)^{1/3}R_*$ its self-gravity will be overcome by the tidal forces of the black hole. 
        As a result, the star will become an elongated stream which is subject to compression during its pericentre passage. 
        At this point, depending on the properties of the black hole and pericentre distance, the stream can precess. 
        \cite{guillochon2015} showed that precession can cause a delayed or prompt triggering of the event as the stream might miss its self-collision during its first turn around the black hole. 
        Following a post-Newtonian approach for the orbital motion and an analytical model to estimate the properties of the stream, \cite{guillochon2015} investigated how streams could wrap around black holes until their self-intersection is achieved. 
        Specifically, they calculated a set of elliptical trajectories that precess at pericentre passage according to post-Newtonian to simulate the motion of the tidal stream. 
        The geometry of the stream is assumed to have a circular cross section whose radius increases each times it goes through pericentre. 
        The calculation is performed until the the trajectory intersect itself,  taking into account the width of the stream. 
        The input parameters to describe this outcome were the following: the star and black hole properties: $m_*$, $R_*$, $M_\text{h}$, $a_\text{h}$, the pericentre distance $r_\text{p}$ (or depth $\beta=r_\text{t}/r_\text{p}$), orbital eccentricity $e$ and misalignment $i$ between the stellar orbit and black hole spin. 

        \begin{figure*}
            \centering
            \includegraphics[width=\linewidth]{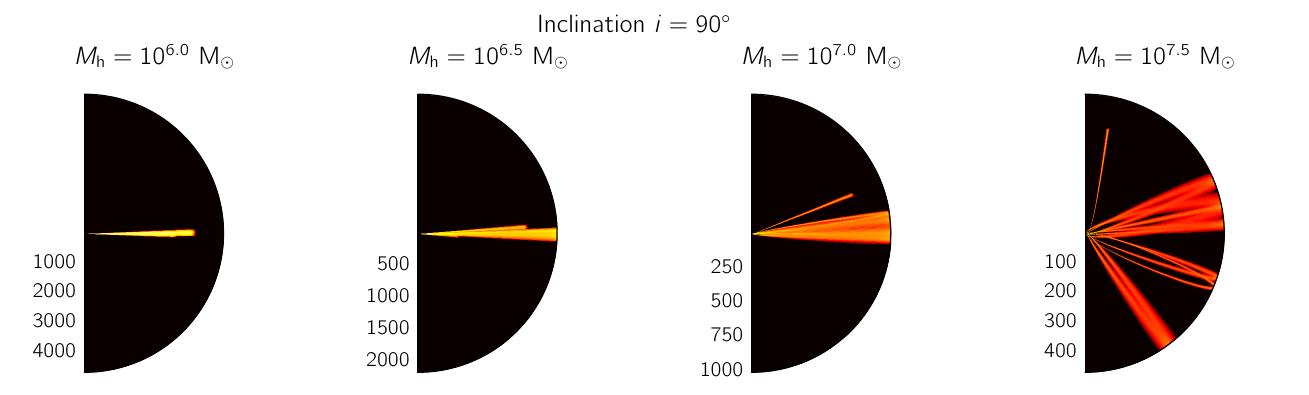}
            \includegraphics[width=0.5\linewidth]{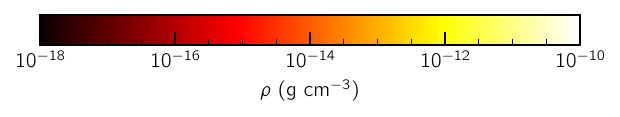}
            \caption{
            Initial conditions for the precessing TDE setup. 
            Each panel shows a two-dimensional $(r,\theta)$ density maps for models with $a_\text{h}=0.9$ for inclination $i=90^{\circ}$ and black hole mass $M_\text{h}=10^{6.0}$, $10^{6.5}$, $10^{7.0}$, $10^{7.5}~\text{M}_{\odot}$. 
            Spatial scales are shown in units of the Schwarzschild radius $R_\text{Sch}=2GM_{\rm h}/c^2$. 
            }
            \label{fig:ics}
        \end{figure*}

    \subsection{Radiation hydrodynamic modelling}
    \label{sec:rhd}

        In this context, we perform two-dimensional $(r,\theta)$ moving-mesh radiation-hydrodynamic simulations using the code JET \citep{duffell2011,duffell2013} with its radiation treatment and coupling module \citep{calderon2021}. 
        The code solves the grey radiation-hydrodynamic equations in the mixed-frame formulation \citep{krumholz2007} in spherical coordinates. 
        The radiation is treated under the flux-limited diffusion approximation \citep{alme1973}.  
        The fluid was considered as an ideal gas with an adiabatic equation of state that is fully ionised and with Solar composition.
        Additionally, we set the sources of opacities using analytical expressions for molecular, H$^-$, electron scattering, and Kramer opacities \citep[e.g.][]{pejcha2017,matsumoto2022}. 

        \begin{figure*}
            \centering
            \includegraphics[width=0.495\linewidth]{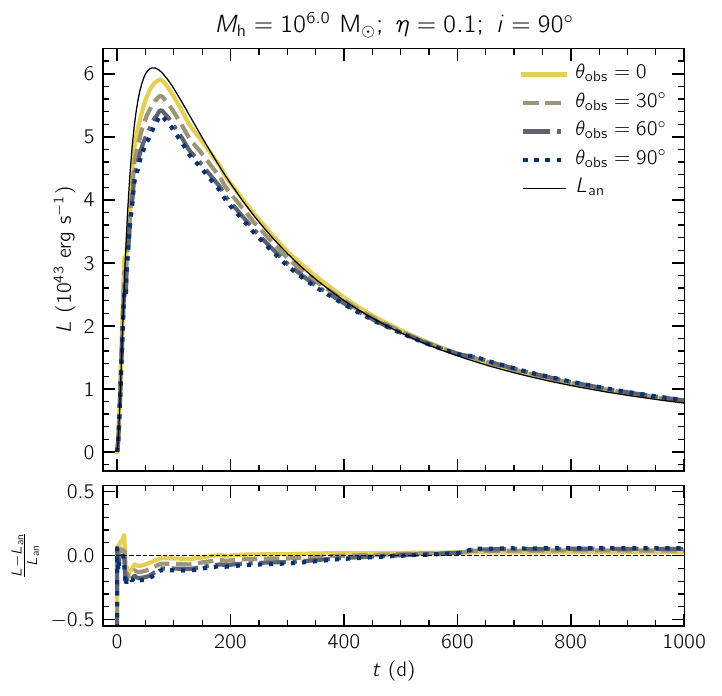}
            \includegraphics[width=0.495\linewidth]{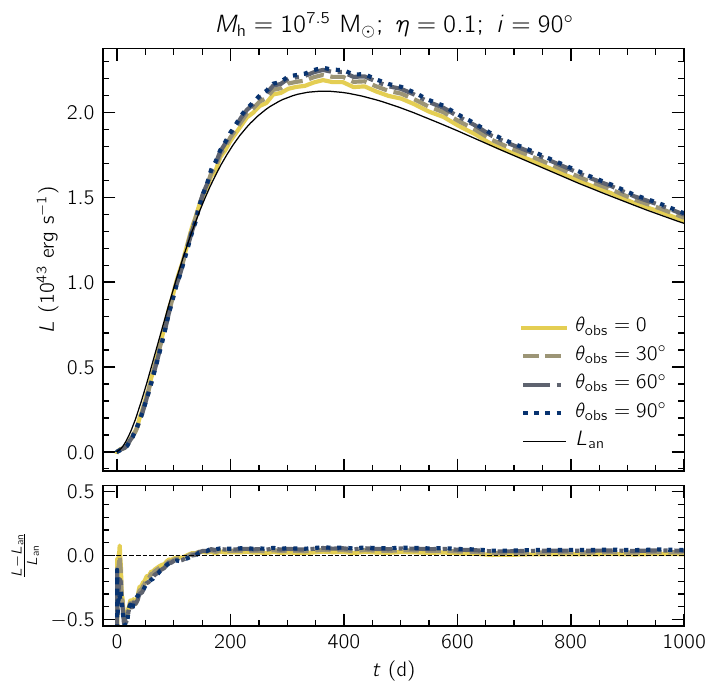}
            \caption{Simulated light curves for models with $M_\text{h}=10^6$ (left-hand side) and $10^7$~M$_{\odot}$ (right-hand side). 
            Each observer line of sight along $\theta_\text{obs}=0,~30^{\circ},~60^{\circ},~90^{\circ}$ is represented with solid yellow, dashed beige, dotted-dashed grey, and dotted blue lines, respectively. 
            The solid thin black line is shown as a reference for an event without surrounding medium \citep{piro2020}. 
            The lower part of each panel shows the relative difference between the simulation and the reference.}
            \label{fig:light_curves}
        \end{figure*}

    \subsection{Numerical setup}
    \label{sec:ics}

        The setup considers a two-dimensional spherical domain with radial and polar extensions: $(r_\text{t},100r_\text{t})$ and $(0,180^{\circ})$, respectively. 
        It is relevant to remark that the domain is allowed to move along the radial dimension based on the moving-mesh capability of JET. 
        The initial density distribution is set to describe the precessed tidal stream in three dimensions, according to the description given in Section~\ref{sec:model}. 
        We set the density in the stream assigning mass proportionally to the local free-fall timescale. 
        We normalise the total mass in the stream so that it is equivalent to  $0.5(1-f_\text{m})m_*$
        This expression comes from assuming that half of the stellar mass remains bound after the disruption, and from this a fraction $f_\text{m}$ is launched in an outflow.
        Finally, it is necessary to integrate the three-dimensional density field along the azimuthal direction since the simulations are in two dimensions $(r,\theta)$.
        Examples of the result of this procedure are shown in Figure~\ref{fig:ics}. 
        This shows density maps for cases with $M_\text{h}=10^6,$ $10^{6.5}$, $10^7$, and $10^{7.5}$~M$_{\odot}$ (from left- to righ-hand side, respectively) while keeping fixed $a_\text{h}=0.9$, $e=0.99$, $i=90^{\circ}$, $\beta=1$, $m_*=1$~M$_{\odot}$, and $R_*=1$~R$_{\odot}$.
        The stream is considered at rest with at a relative low temperature ($T=10^3~\text{K}$) and low radiation energy density ($E_\text{r}=10^{-16}~\text{erg}~\text{cm}^{-3}$). 
        The ``empty" regions of the domain are filled with material with low density, thermal pressure, and radiation energy density for numerical reasons. 

        The outflow and luminosity of the TDE are included as time-dependent boundary conditions at the innermost edge of the domain. 
        The outflow is modelled as a spherically symmetric wind whose mass-loss rate is set to the mass fallback rate of a TDE $M_\text{fb}(t)$. 
        This rate is calculated considering the disruption of a stellar polytrope with $n=1.5$ \citep{lodato2009}. 
        The total mass in the outflow corresponds to $f_\text{m}$ of the bound material of the star. 
        The outflow is launched at the minimum speed to unbind material at a distance $r_\text{t}$.
        The temperature and radiation energy density of the wind were set to small values so that they do not have an impact on its dynamics. 
        The injected luminosity is estimated from the assumption that only a small fraction $f_\text{in}$ of the bound material is accreted.  
        Then, $L(t)=\eta f_\text{in}f_\text{m}M_\text{fb}c^2$, where $\eta$ is the efficiency parameter. 
        The simulations were run for a total of 1000 days. 

\section{Results}
\label{sec:results}

    Here we present the results for the models with $M_\text{h}=10^6~\text{M}_{\odot}$ and $10^{7.5}~\text{M}_{\odot}$ while the rest of the parameters were kept set to $\beta=1$, $M_*=1$~M$_{\odot}$, $R_*=1$~R$_{\odot}$, $e=0.99$, $i=90^{\circ}$, and $\eta=0.1$. 
    Notice that the initial conditions of these models correspond to the left- and right-hand side panels in Figure~\ref{fig:ics}. 

    We synthesised light curves along different lines of sight. 
    To do so, we mapped the two-dimensional domain in three dimensions, assuming azimuthal symmetry. 
    Afterwards, we searched for the location of the photosphere $r_\text{ph}$ along radial rays: $\tau(r=r_\text{ph})=2/3$. 
    Then, we projected the radiative flux at the location of the photosphere along an arbitrary line of sight $\theta_\text{obs}$. 
    Finally, we integrate the resulting quantity over the photosphere surface.
    Repeating this procedure on every snapshot and for different lines of sight we obtain a set of light curves for each simulation. 
    Figure~\ref{fig:light_curves} shows the light curves for the two models and lines of sight $\theta_\text{obs}=0$ (polar direction) $,30^{\circ}$, $60^{\circ}$, and $90^{\circ}$ (equatorial direction). 
    As a reference, a semi-analytical model of reprocessed emission of the injected luminosity and wind is shown as a solid black line \citep{piro2020}. 
    The left- and right-hand side panels show the events with $M_\text{h}=10^6$~M$_{\odot}$ and $10^{7.5}$~M$_{\odot}$, respectively. 
    In both cases there is extra attenuation along all lines of sight when compared with the reference model. 
    The degree of attenuation is about 10-20\% in the case of a less massive black hole being the most along the stream since it is optically thick. 
    Although for a more massive black hole the extra attenuation can reach $\sim$50\% this seems to affect to the same extent to all lines of sight. 
    This causes a delay in reaching the peak of 100-200 days. 
    The different degree of impact of the stream on the light curves can be explained by the fact that changing the mass of the black hole affects the scale of the system. 
    That is, keeping $\beta$ constant the pericentre is located further for a more massive black hole. 
    As a result, the stream mass is distributed over a larger volume, and the precession out of the orbital plane adds to this effect. 
    Then, the significant attenuation occurs only on short timescales and the source looks isotropic on 100~days. 
    In the case of the smaller black hole, the stream is denser but affects significantly only lines of sight along its orientation, and the source looks isotropic on slightly longer timescales ($\gtrsim200$~d). 
    For an in-depth discussion of the impact of the other input parameters we refer the reader to \cite{calderon2024}. 
    The models studied here covered only completely disrupted stars. 
    In the case of partial tidal disruption events, it is expected that the luminosity and the wind are weaker but also the total bound material would be less. 
    Thus, it is not obvious to expect more relevant signatures due to precession in the light curves of such events.

    \begin{figure*}
        \centering
        \includegraphics[width=0.85\linewidth]{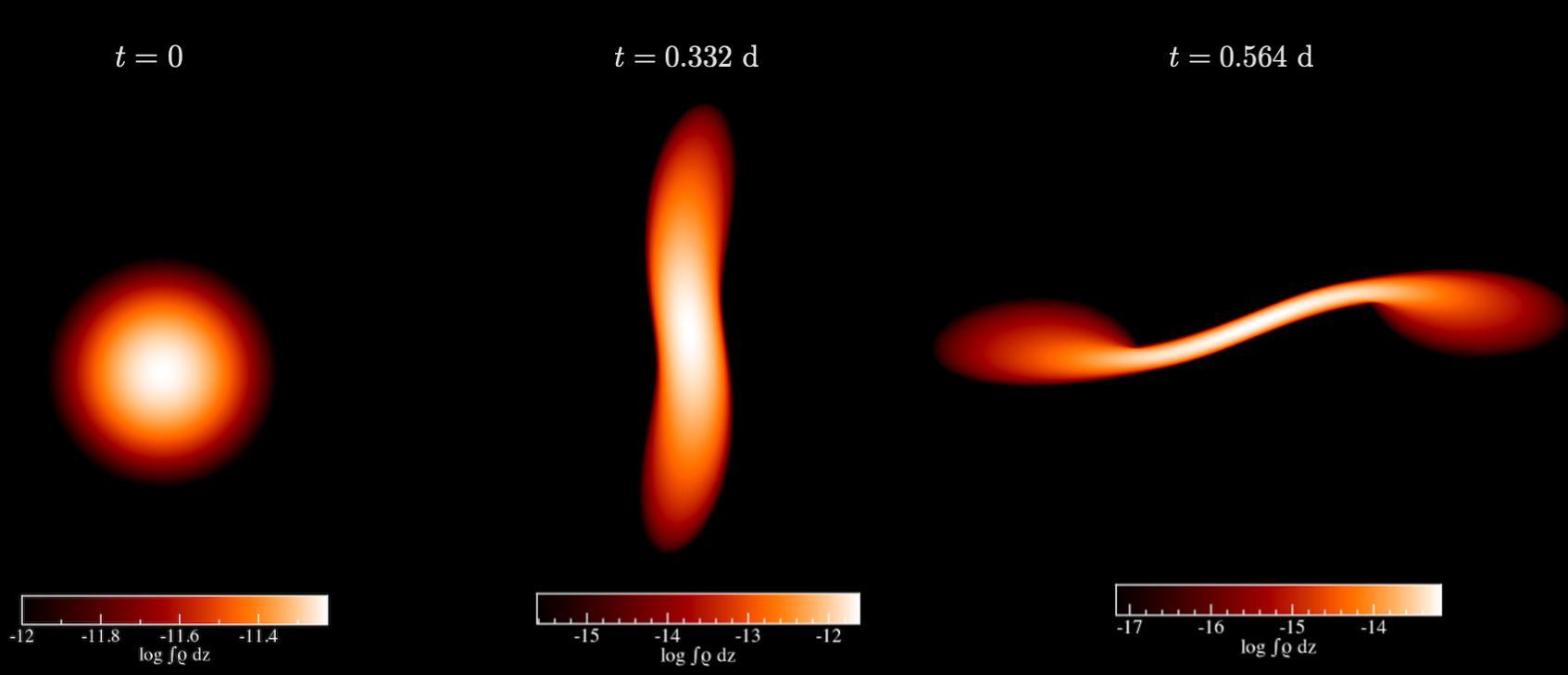}
        \caption{
        Stellar disruption evolution during pericentre passage from the simulation with SPHINCS. 
        The panels correspond to density maps integrated along the $z$ axis that is perpendicular to the orbital plane. 
        These maps were made using the code splash \citep{price2007}. 
        Notice that every panel does not show the same spatial scale.
        }
        \label{fig:snaps}
    \end{figure*}

\section{Towards more realistic initial conditions}
\label{sec:sph}

\begin{figure}
        \centering
        \includegraphics[width=\linewidth]{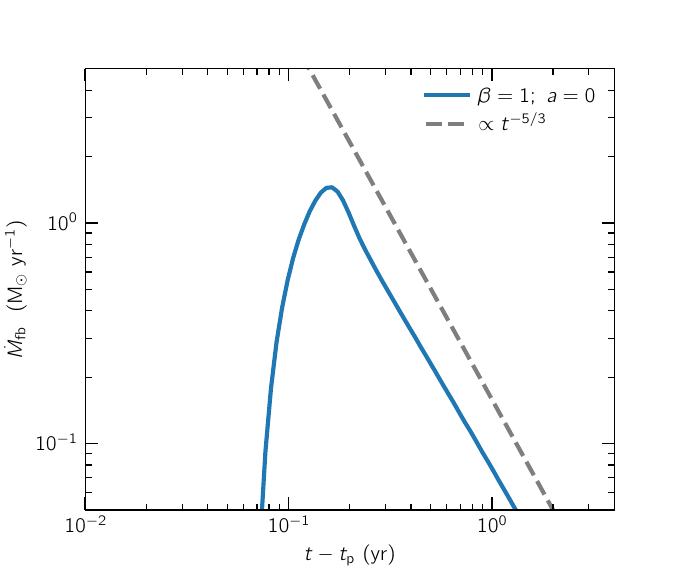}
        \caption{
        Mass fallback rate as a function of time from the first pericentre passage (solid blue line).  
        The dashed gray line represents the $t^{-5/3}$ decay. 
        }
        \label{fig:fallback}
    \end{figure}

    The initial conditions of these models were based on analytical geometric and post-Newtonian precession approaches. 
    The scenario of a disrupted stellar stream wrapping around a spinning black hole until its own self-collision is in itself a challenging problem. 
    In this context, we have adapted the code SPHINCS \citep[``Smoothed-Particle Hydrodynamics in Curved Spacetime";][]{rosswog2021,rosswog2023}, so that is capable to model stellar disruptions due to the presence of rotating black holes. 
    The code is designed to self-consistently evolve dynamically changing space-times but it can also be used to evolve gas flows in Kerr metric.
    In the case of TDEs, the mass ratio is usually very small ($q=m_*/M_\text{h}\ll1$), and thus it is reasonable to consider a time-independent background metric determined by the central (spinning) black hole.  
    The exact implementation and validation tests will be presented in a dedicated article (Calderón \& Rosswog, in prep.). 
    Here, we present an example of the capabilities of our implementation to model a Solar-like star ($m_*=1$~M$_{\odot}$; $R_*=1$~R$_{\odot}$) on a parabolic orbit launched at $10r_\text{t}$ towards a black hole ($M_\text{h}=10^6$~M$_{\odot}$; $a_\text{h}=0$) with $\beta=1$ and $i=0$. 
    The star was modelled as a polytrope with $n=3$ and previously evolved in a flat spacetime so that it is only subject to hydrodynamics and self-gravity interactions until it reaches an equilibrium state. 
    The star was modelled using 4,194,304 particles and evolved for 30~days of simulation time. 
    Figure~\ref{fig:snaps} present three snapshots of the integrated density along the axis perpendicular to the orbital plane to see the evolution of the stellar disruption during its pericentre passage. 
    As a consistency test, we have calculated the mass fallback rate for this fiducial test following the method used by \cite{gafton2019}.
    We extracted the properties of all bound particles in the final snapshot and computed the time coordinate until the next pericentre passage. 
    The result was binned into a histogram and we assigned weights corresponding to the mass of each SPH particle divided over the time bin width. 
    This procedure outputs the mass fallback rate computed directly from the simulation under the general relativistic framework. 
    Figure~\ref{fig:fallback} shows the mass fallback rate as a function of time extracted from the simulation. 
    Notice that this agrees well with the expected rapid rise to its peak and decay following the expected theoretical power law. 
    This result was contrasted with analogous calculations performed by \cite{gafton2019}. 
    The comparison shows good agreement since both fallback mass rate peaks are $\sim3$~M$_{\odot}~$~yr$^{-1}$ at $\sim$60~d from pericentre passage, and the power law decays also follow -$5/3$ \citep[see Figure 8, 9, and 10 in][]{gafton2019}. 
    Simulating this fiducial case appropriately either for longer ($t\gtrsim1$~yr) or other deeper encounters ($\beta>1$) is not straightforward due to the need of significantly higher resolution \citep[$\gg10^8$ particles;][]{bonnerot2020}  which in some cases can be computationally prohibited. 
    State-of-the-art works have opted for developing for alternatives techniques to overcome this issue such as simulating the star in a box moving along its orbit and then remapping it into a larger domain to follow the longer term evolution \citep{ryu2023b}, or using adaptive particle refinement in the SPH approach \citep{hu2025}, among others.

\section{Summary}
\label{sec:summary}

    We have conducted two-dimensional radiation-hydrodynamic simulations of the interaction between the TDE radiation and outflow with the surrounding structure expected to result from relativistic precession. 
    We have explored the impact of the black hole mass and misalignment between the stellar orbit and black hole spin. 
    We have synthesised light curves for observers along different lines of sight to identify potential signatures imprinted in them. 
    The results showed that the effects of the interaction is overall mild and depends on the line of sight. 
    For events with black holes of 10$^6$~M$_{\odot}$, the light curves along the direction of the stream show a higher degree of obscuration with respect to the fiducial case. 
    However, the source becomes isotropic on timescales of $\sim$100~days. 
    For events with high degree of misalignment and black holes of 10$^{7.5}$~M$_{\odot}$, the light curves show an extra delay of 100-200 days on reaching its peak along all lines of sight. 
    Although these scenarios affect the generic TDE light curve it might not be possible to identify when having only a single line of sight of the event. 

    In this work, we have also shown our efforts to calculate more realistic initial conditions adapting the general relativistic SPH code SPHINCS to model stellar disruptions. 
    Currently, the code can simulate stellar disruptions in curved spacetime for rotating black holes and recover successfully the expected mass fallback rates. 
    However, it remains as a challenge to use it to simulate the year-long evolution of TDE or deeper events.

\section*{Acknowledgments}

    This work was performed under the auspices of the \fundingAgency{Alexander von Humboldt Foundation}. 
    The research of D.C. and S.R. has been funded by the \fundingAgency{Deutsche Forschungsgemeinschaft} (DFG, German Research Foundation) under Germany’s Excellence Strategy - EXC 2121 - ``Quantum Universe” - \fundingNumber{390833306}. 
    O.P. has been supported by the \fundingAgency{Czech Science Foundation} under grant number \fundingNumber{24–11023S}.
    S.R. has been supported by the \fundingAgency{Swedish Research Council} (VR) under grant number \fundingNumber{2020-05044}, by the research environment grant ``Gravitational Radiation and Electromagnetic Astrophysical Transients” (GREAT) funded by the Swedish Research Council (VR) under Dnr \fundingNumber{2016-06012}, by the \fundingAgency{Knut and Alice Wallenberg Foundation} under grant Dnr. KAW \fundingNumber{2019.0112}, and by the \fundingAgency{European Research Council} Advanced Grant `INSPIRATION' under grant agreement No. \fundingNumber{101053985}. 
    This work was supported by the Ministry of Education, Youth and Sports of the Czech Republic through the e-INFRA CZ (ID:90140 and ID:90254).
    Part of this work was developed at the Aspen Center for Physics, which is supported by the \fundingAgency{National Science Foundation of the USA} under project No. \fundingNumber{PHY-1607611}.
    In addition, D.~C. acknowledges the kind hospitality of the Center for Computational Astrophysics, where part of this project was conducted.
    This work made use of \textsc{python} libraries \textsc{numpy} \citep{harris2020} and \textsc{matplotlib} \citep{hunter2007}, as well as of the NASA’s Astrophysics Data System.

\subsection*{Author contributions}

    D.~C. contributed adapting the numerical tool, setting up, running, and analysing the numerical simulations. 
    He wrote the article, reviewed and edited it.
    O.~P. contributed with some of the main ideas and discussions of the project. 
    He also reviewed and edited the article.
    B.~D.~M. contributed with the conceptualisation, main foundations and definition of the problem. 
    He also reviewed and edited the article.
    P.~C.~D. contributed developing and maintaining the numerical tool JET, as well as reviewing and editing the article. 
    S.~R. is the main developer of the general relativistic hydrodynamic code SPHINCS, assisted with the setup and numerical simulations.
    
\subsection*{Financial disclosure}

    None reported.

\subsection*{Conflict of interest}

    The authors declare no potential conflict of interests.



\bibliography{Wiley-ASNA}%



\end{document}